# Photoinduced electromotive force on the surface of InN epitaxial layers

B. K. Barick and S. Dhar*

*Department of Physics, Indian Institute of Technology, Bombay, Mumbai-400076, India*

*Corresponding Author's Email: dhar@phy.iitb.ac.in

## Abstract:

We report the generation of photo-induced electromotive force (EMF) on the surface of c-axis oriented InN epitaxial films grown on sapphire substrates. It has been found that under the illumination of above band gap light, EMFs of different magnitudes and polarities are developed on different parts of the surface of these layers. The effect is not the same as the surface photovoltaic or Dember potential effects, both of which result in the development of EMF across the layer thickness, not between different contacts on the surface. These layers are also found to show negative photoconductivity effect. Interplay between surface photo-EMF and negative photoconductivity result in a unique scenario, where the magnitude as well as the sign of the photo-induced change in conductivity become bias dependent. A theoretical model is developed, where both the effects are attributed to the 2D electron gas (2DEG) channel formed just below the film surface as a result of the transfer of electrons from certain donor-like-surfacestates, which are likely to be resulting due to the adsorption of certain groups/adatoms on the film surface. In the model, the photo-EMF effect is explained in terms of a spatially inhomogeneous distribution of these groups/adatoms over the surface resulting in a lateral non-uniformity in the depth distribution of the potential profile confining the 2DEG. Existence of such an inhomogeneity in the distribution of surface potential has indeed been experimentally found for these layers.

# I. Introduction

Indium nitrides (InN) is the least understood material among the three of the group III-nitride family, where GaN and AlN are the other two members. Until recently, there was a big controversy over the band gap of this material. The band gap was earlier believed to be about 1.9 eV [1] but recent studies assign a much lower value of 0.65 eV to it[2]. This makes this material highly attractive for infrared optoelectronics, such as photo-diode [3], detectors [4] as well as for solar cell applications [5]. Among all group III-nitrides, InN has the lowest electron effective mass. [6] This is the reason why the highest mobility reported in this material is several times more than those reported for GaN and AlN.[7] However, the growth of high quality InN is still a challenge. InN layers are often found to be degenerate with high background electron concentrations resulting from unintentional doping. [8] On the other hand, this degenerate nature is believed to be responsible for certain fascinating phenomena such as negative photoconductivity (NPC) [9] and the transition to superconductivity at low temperatures [10] observed in this material. NPC has been observed in doped narrow band gap semiconductors such as GaAs [11], $In_2Se_3$ [12], InAs nanowires [13] and heterojunctions such as p-InSb/i-GaAs [14], GaAs/AlGaAs quantum wells [15]. Wei et al. [9] have reported the NPC in degenerate InN layers. They explained the phenomenon in terms of a reduction of the carrier mobility upon photoexcitation. According to the model, photoexcited holes are trapped in certain defect/impurity levels lying close to the valance band, this increases the ionized impurity scattering rate of the carriers, resulting in a reduction of mobility and hence the overall conductivity. Guo et al. [16] have reported a transition from negative to positive photoconductivity in InN as the temperature is decreased from 300 to 100K. Furthermore, the transition temperature is found to depend inversely on the carrier concentration. The same group have also observed a negative to positive photoconductivity transition with the increase of Mg doping in InN.[17] This effect is attributed to the transition from n-type to p-type conduction in the material.[17]

Another interesting aspect of InN epitaxial layers is the bending of the band, which is associated with the accumulation of electrons just below the surface. This thin two dimensional electron gas (2DEG) channel can have a much higher mobility as compared to the underlying bulk layer as a result of the suppression of scattering cross-section due to reduced dimensionality and hence can dominate the conducting property of the material. Surface is thus expected to play a crucial role in governing any photoinduced change in the conduction property of InN layer. For example, attachment of certain groups/adatoms on the surface can result in a distribution of energy levels in the forbidden gap. These states can play a crucial role in governing the redistribution of the photo-generated carriers in the 2DEG region as these carriers can easily be transferred to/from these states.

Here, we report the development of electromotive force (EMF) on the surface of InN epitaxial films when illuminated with above band gap light. In this effect, with respect to a fixed point on the surface, different parts are found to deliver EMF of different magnitudes and polarities. Notably, the effect is not the same as the surface photovoltaic effect [18] or Dember potential [19], in both of which EMF is developed across the layer thickness, not between different contacts on the surface. Furthermore, these layers are found to show negative photoconductivity effect. Coexistence of surface photo-EMF and negative photoconductivity effects result in a unique photoconductivity response, in which not only the magnitude but also the sign of the photoconductivity depend on the applied bias. A theoretical model is developed, where it is assumed that the attachment of certain groups/adatoms on the surface results in the formation of donor-like-surfacestates. In thermal equilibrium, transfer of electrons from these states to the bulk of the material leads to the formation of a thin layer of 2D electron gas (2DEG) just below the surface. A spatial inhomogeneity of the attachment of these groups/adatoms with the surface can give rise to an inhomogeneity in the surface band bending and hence in the depth of the electron confining potential (surface potential) leading to a non-uniform distribution of drift and diffusion current densities over the 2DEG channel. In thermal equilibrium, the drift and diffusion current balance each other in a way that the net current density maintained to be zero everywhere. However, this condition is violated upon illumination, when a net lateral current density distribution could be set in the 2DEG region. A photo-induced EMF can thus be developed between any two points on the surface. The magnitude and the orientation of the EMF are governed solely by the spatial variation of the surface potential between the two points. Kelvin probe force microscopy studies on these layers indeed show the existence of a non-uniform distribution of the surface potential.

## II. Experimental Details:

c-oriented InN epitaxial films of thickness ~200 nm were grown on c-sapphire substrates by chemical vapour deposition (CVD) technique. More details about the growth, structural and electronic properties of these films can be found elsewhere [20]. See the supplementary information for the basic structural, morphological, band gap and electrical properties of the InN epitaxial layers studied here. XPS measurements were carried out using MULTILAB from Thermo VG Scientific using Al K-α as x-ray source (1486.6 eV). Indium contact pads were fabricated at the four corners of squire shaped samples. These contacts were found to show ohmic behaviour. Samples were mounted on the cold finger of an optical window fitted liquid nitrogen cryostat. A temperature controller (Cryocon 32B) was used to stabilize and measure the temperature at the sample location. In photo-EMF measurements, voltage between indium contacts was recorded using a nanovoltmeter when the layer is illuminated with an above band gap light for which either a Xe light source or a 515 nm diode laser were used. These measurements were carried out under various environmental conditions such as vacuum, normal atmospheric condition and humidity. In order to create humid condition surrounding the sample, argon gas was passed through a water column before introducing it into the measurement enclosure. Data were recorded at various temperatures and photo-excitation powers. In photoconductivity measurements, photo-induced change in conductance was measured at different applied voltages using a Keithely made picoammeter-voltage source. Kelvin probe force microscopy (KPFM) were carried out using a MFP3D Origin AFM system from Asylum/Oxford instruments. Images were recorded on different parts of the sample surface under dark and illuminated conditions.

## III. Results and discussions:

### A. Surface photo-EMF

Figure 1(a) shows the time response of the photovoltage ($V_{ph}$) measured between different contacts fabricated on the surface of an InN layer as it is exposed to white light for some time and then the exposure is switched off. Indium contacts at the four corners of the squire shaped sample are schematically shown in the inset of the figure. Evidently, when the sample surface is exposed to white light, $V_{ph}$ increases to a saturation ($V_{ph}^{max}$) in a timescale of several tens of seconds. The photovoltage returns to zero in a similar timescale after switching the excitation off. Interestingly, the polarity of $V_{ph}$ is different for different contact pairs. For example, potential with respect to contact 2 is found to be positive at contact 3, while it is negative at 1 and 4. Fig. 1(b) compares the time variation of the photovoltage developed between contacts 2 and 3 at different environmental conditions. Upon evacuation, the saturation value of the photovoltage is found to be more than that is recorded in ambient condition. The magnitude of $V_{ph}^{max}$ decreases further under humidity. These observations clearly suggest that the effect must be associated with the surface. Moreover, there must be some connection between the inhomogeneity of the surface and the observation of different polarity of the photovoltage at different contacts.

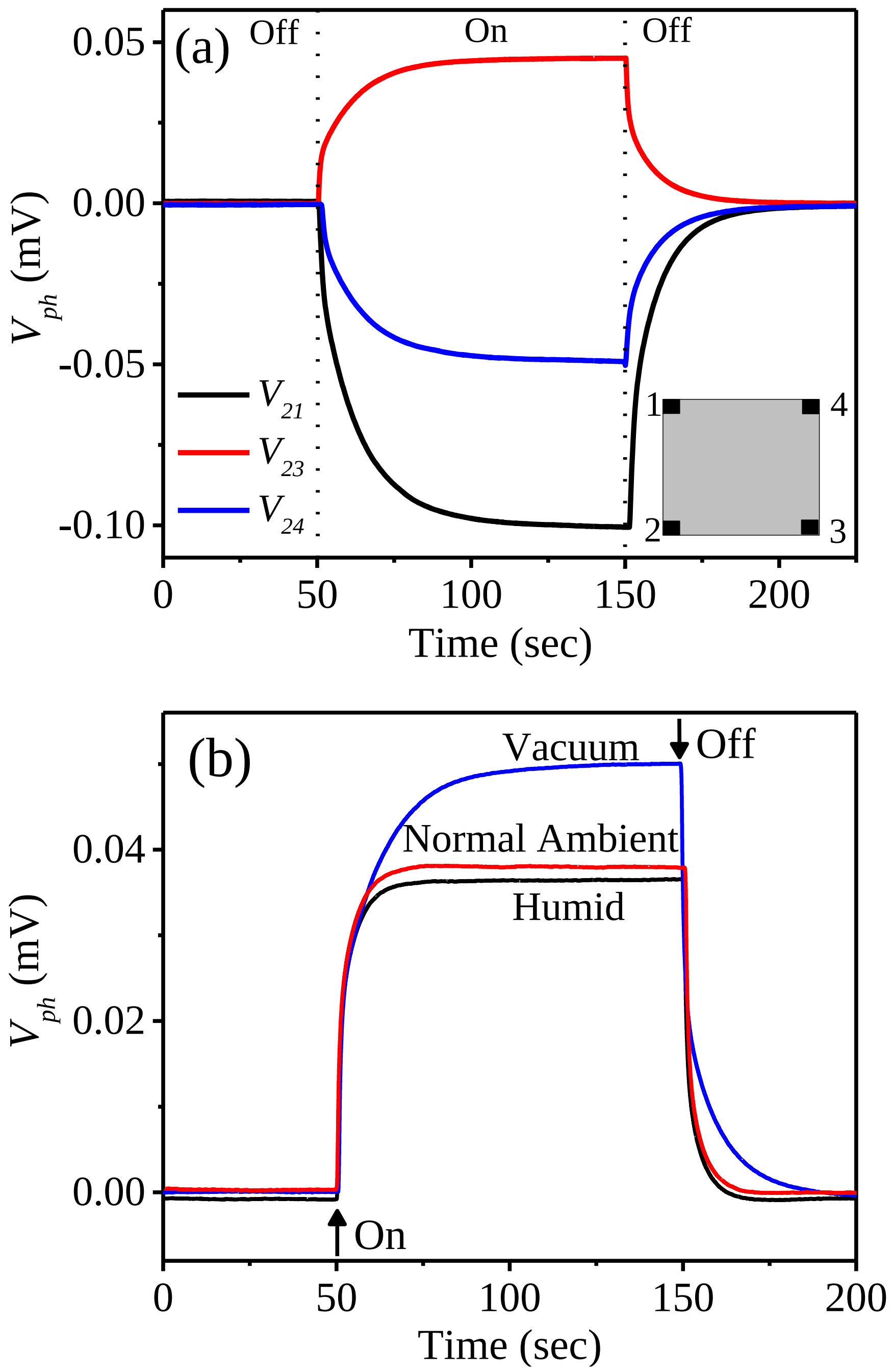

Fig. 1. (a) Time dynamics of the photovoltage ( $V_{ph}$ ) generated between different contacts fabricated on the surface of an InN layer as it is exposed to white light for some time and then the exposure is switched off. Inset schematically depicts Indium contacts at the four corners of the squire shaped sample. (b) Time profiles for the photovoltages developed between contacts 2 and 3 at different environmental conditions.

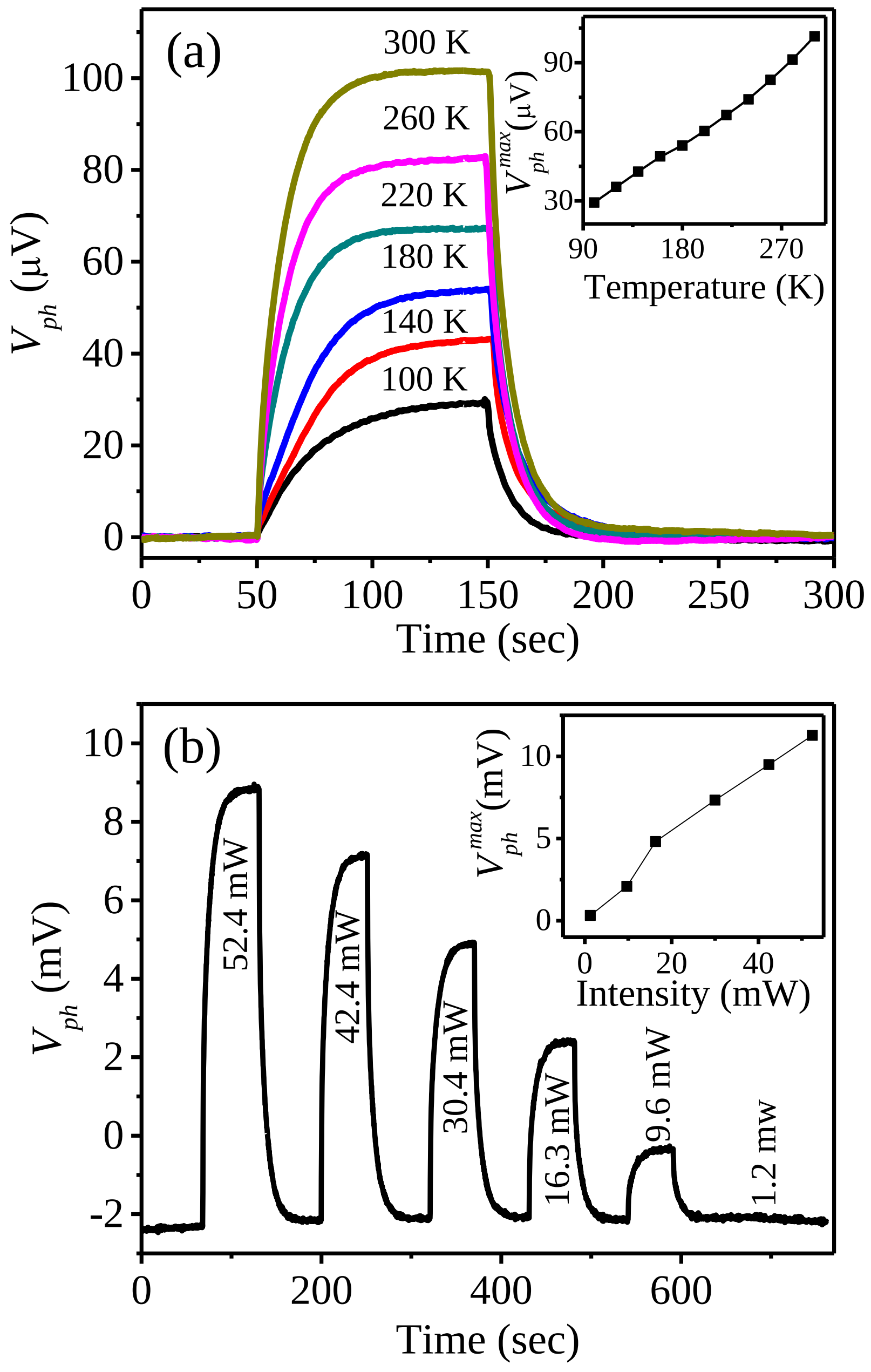

Fig. 2. (a) Time variation of the photovoltage response recorded between two contacts at different temperatures. Inset shows the magnitude of the photovoltage build-up $V_{ph}^{\max}$ as a function of temperature. (b) Time variation of $V_{ph}$ recorded between the contacts as the sample is illuminated intermittently with a 515 nm laser light of different intensities. Insert shows $V_{ph}^{\max}$ as a function of excitation intensity.

Figure 2(a) represents the time variation of the photovoltage response recorded between two contacts at different temperatures. It is evident from the figure that both the growth and decay times for the photo-response increase with the reduction of temperature. Moreover, the magnitude of the photovoltage build-up $V_{ph}^{\max}$ steadily decreases with temperature as shown in the inset of the figure. Fig. 2(b) shows the time variation of $V_{ph}$ recorded between the contacts as the sample is illuminated intermittently with a 515 nm laser light of different intensities. It is noticeable that $V_{ph}^{\max}$ increases with the excitation power. This variation is plotted in the inset of the figure.

## B. Negative Photoconductivity

Figure 3 compares the photo-conductance response profiles recorded under different applied voltages between contacts located at different parts of the sample surface. When the applied bias is sufficiently high, in both the cases, conductance decreases upon illumination, exhibiting a negative photoconductivity effect. As the applied bias $V_b$ decreases, negative photo-conductance $G_{ph}$ turns positive for the case of Fig. 3(a). While, $G_{ph}$ remains to be negative for the other contact pair [Fig.3 (b)]. Let us consider that the resistance before illumination between a pair of contacts is $R_o$. It is increased to $R_o + R_{ph}$ upon illumination due to negative photoconductivity effect. Now the photo-conductance can be expressed as $G_{ph} = (V_b + V_{ph})/(R_o + R_{ph})V_b - 1/R_o = (R_o V_{ph} - R_{ph} V_b)/R_o(R_o + R_{ph})V_b$. Clearly, $G_{ph} \approx -R_{ph}/R_o(R_o + R_{ph})$ is negative for sufficiently high $V_b$ satisfying $R_{ph}|V_b| >> R_o|V_{ph}|$, irrespective of the polarities of $V_b$ and $V_{ph}$. On the other hand, $G_{ph} \approx R_o V_{ph}/R_o(R_o + R_{ph})V_b$ for $R_{ph}|V_b| << R_o|V_{ph}|$. $G_{ph}$ can thus be either negative or positive depending on whether the polarities of $V_b$ and $V_{ph}$ are mutually opposite or not. In the first case [Fig. 3(a)] polarities of $V_b$ and $V_{ph}$ must be the same, while they are mutually opposite in the second case [Fig. 3(b)]. These relationships between $V_b$ and $V_{ph}$ are indeed been found to hold in these two pairs of contacts. These results clearly show that both the surface photo-EMF and negative photoconductivity effects, are together active when the sample is illuminated with an above band gap light. It should be noted that negative photo-conductivity has been reported in InN epitaxial layers by a number of groups [9,16,17]. In a few of these reports, a transition from positive to negative photoconductivity has been observed as a function of temperature [16] and doping concentration [17].

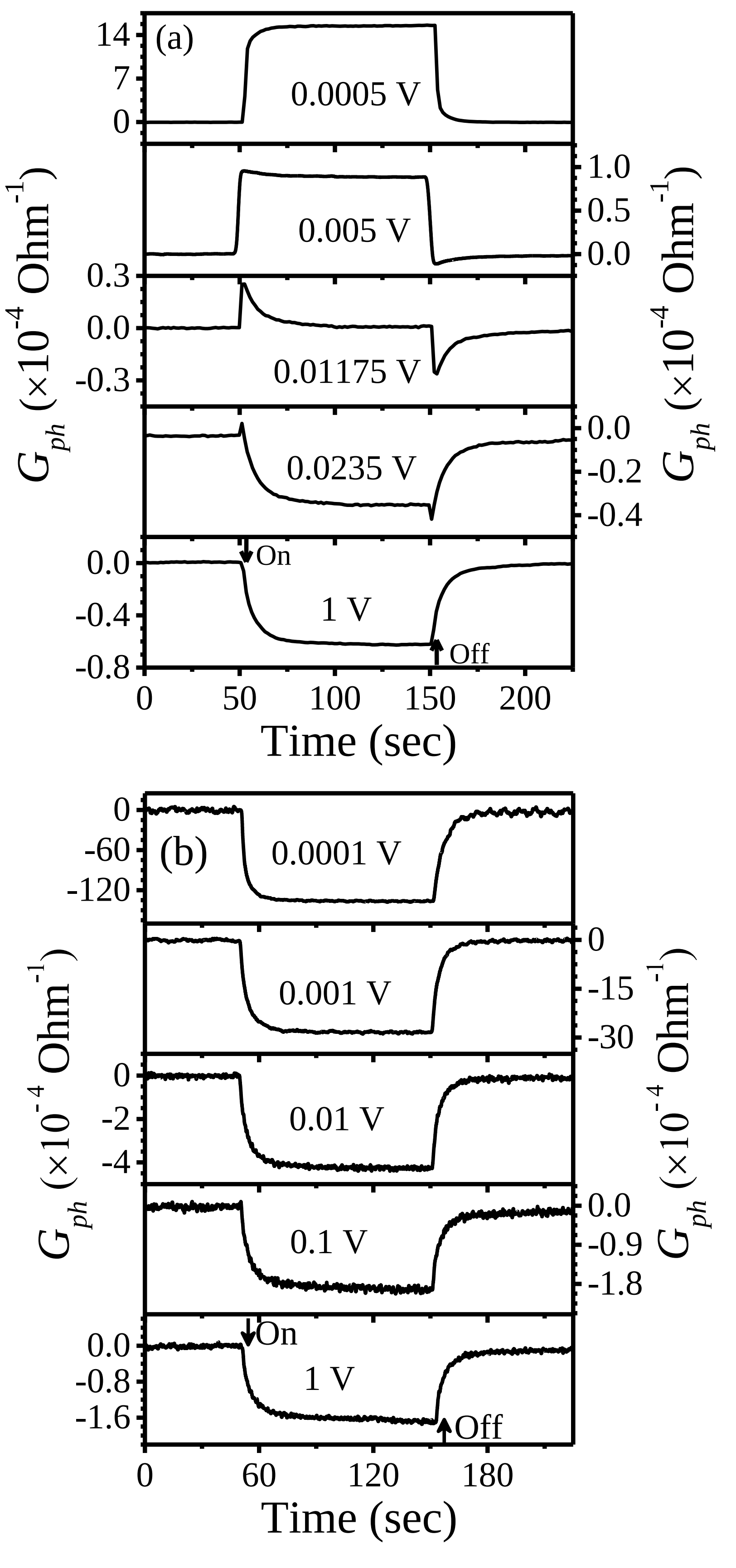

Fig. 3. Photo-conductance response profiles recorded under different applied voltages between contacts located at two different locations of the sample surface.

C. Surface Fermi Level

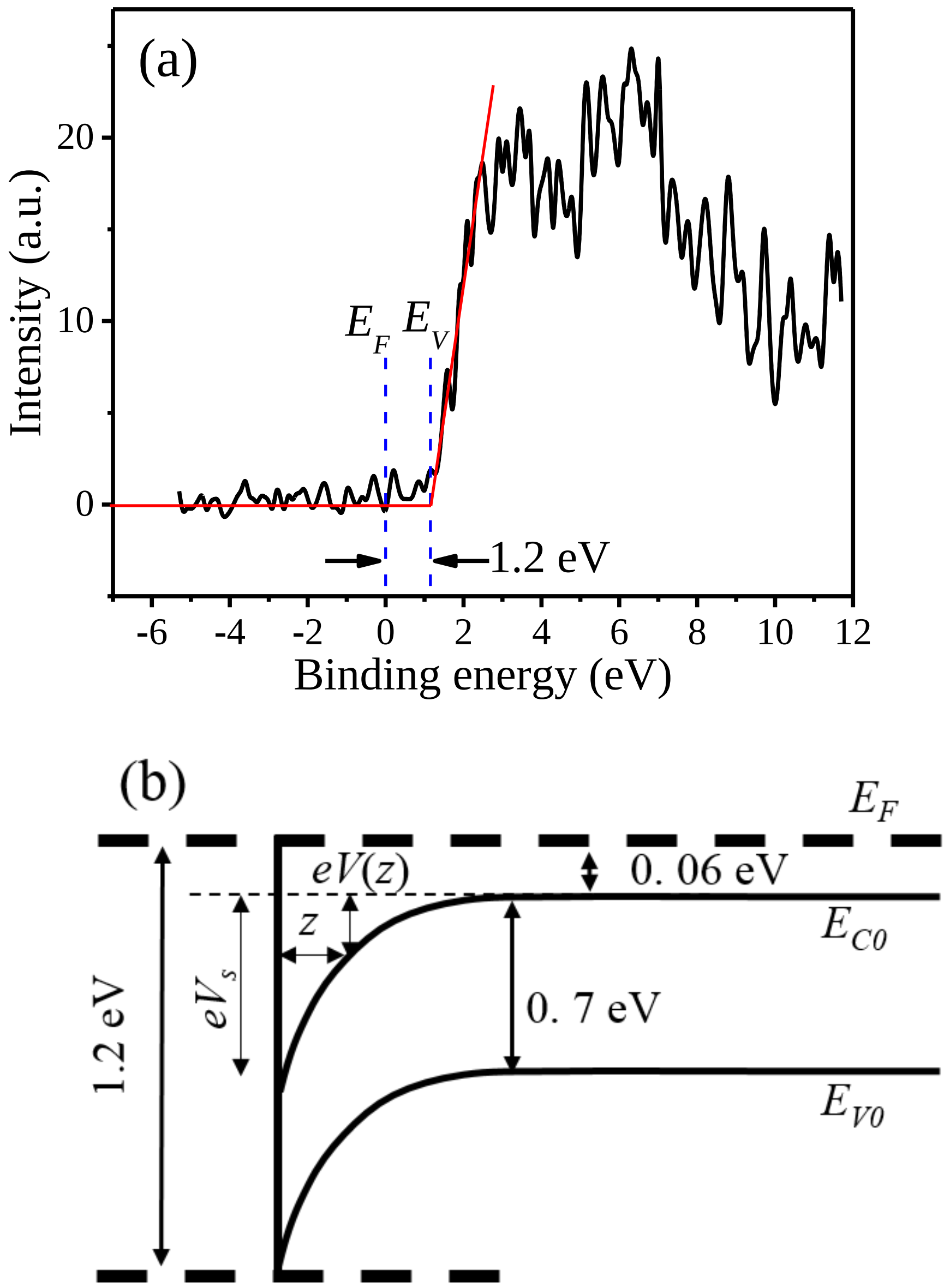

Fig. 4. (a) XPS spectrum recorded for the sample at the valence band edge. (b) Schematic representation of the band bending at the surface of the layer.

Figure 4(a) shows XPS spectrum recorded for the sample at the valence band edge. Note that the zero energy here represents the position of the Fermi level at the surface. Position of the valence band maximum ($E_V$) is estimated to be 1.2 eV below the surface Fermi level ($E_{FS}$). However, InN has a band gap of 0.7 eV at room temperature [2], which means that in this sample, surface Fermi level is located 0.5 eV above the conduction band minimum ($E_C$). Pinning of Fermi level above the conduction band suggests an accumulation of electron at the surface of the film. In fact, the effect of surface accumulation of electrons has often been observed in InN films [21] and nanowires [22]. Note that the Hall measurement reveals an electron concentration of ~ $2\times10^{19}$ cm$^{-3}$ in this sample, which shows that the material is fully degenerate. At this electron concentration, the position of the Fermi level lies 0.06 eV above the conduction band minimum in the bulk of the layer. The origin of such a high background electron concentration in this sample can be attributed to certain unintentional donors. It is noteworthy that high background electron concentration in as-grown epitaxial InN layers has frequently been reported. However, the identity of the unintentional donor is still unclear, as different groups have attributed the background electron concentration to point defects such as N-vacancy [23] and impurities [24, 25]. Variation of the conduction and valence band edges as a function of the distance from the surface is shown schematically in Fig. 4(b), where the downward bending of the band at the surface is associated with the electron accumulation as suggested by the XPS study [Fig. 4(a)].

Note that the photo-EMF effect observed in Fig.1 should not be confused with the surface photovoltaic effect, which arises due to the movement of photo-excited electrons and holes in opposite directions at the surface of a semiconductor, where the band bends as a result of depletion or accumulation of charges. In surface photovoltaic effect, a voltage is thus developed between the surface and the bottom of the layer. In this case, however, photovoltage is developed between different parts of the film surface. Moreover, the magnitude of the voltage depends strongly on the surrounding conditions [Fig. 1(b)], suggesting that the effect is closely connected to the surface. In order to understand the effect, a theoretical model can be set up as follows.

## D. Theoretical Model

We assume that the attachment of certain groups/adatoms on the surface results in the formation of certain donor-like-surfacestates. In thermal equilibrium, transfer of electrons from these donors to the bulk of the material leads to the downward bending of the band at the surface of film in order to accommodate the excess electrons as shown in Fig. 4(b). Quantum confinement of electrons in this triangular quantum well constitute a thin layer of 2D electron gas just below the surface, which we have termed as 2D electron accumulation (2DEA) region. At thermodynamic equilibrium, the density of charge developed at the surface as a result of ionization of these surface donors due to the electron transfer from these surface states to the bulk can be expressed as $Q_{ss} = eN_{sd} / \left[1 + \exp\left\{\left(E_F + E_{sd} + eV_s\right)/ k_B T\right\}\right]$ [18], where $E_{sd}$ is the energy separation between the surface donor level and the conduction band minimum at the surface. While, $N_{sd}$ is the areal density of surface donors. $E_F$ is the fermi level position with respect to the conduction band minimum in the charge neutral region (away from the surface, where the band becomes flat) and $V_s$ is the surface potential [see Fig. 4(b)]. Charge density $\rho(z)$ at a depth of $z$ from the surface of the layer can be expressed as $\rho(z) = e[N_d^+(z) - n_{3D}(z)]$, where $N_d^+(z) = N_d / 1 + 2\exp\left[\{E_F + E_d + eV(z)\}/k_B T\right]$ the volume density of ionized donors at $z$. $N_d$ the density of donors in InN layer. While, $E_d$ the donor level position from the conduction band minimum and $V(z)$ is the potential such that $E_c(z) = E_{co} - eV(z)$, where $E_c(z)$ and $E_{co}$ represent the conduction band minimum at z and in the charge neutral region, respectively. $n_{3D}(z) = N_c F_{1/2}[\{E_F + eV(z)\}/k_B T]$, where $N_c = 2\left(m_e^* k_B T / 2\pi\hbar^2\right)^{3/2}$. $F_{1/2}(\eta)$ the Fermi integral and $m_e^*$ the electron effective mass in the conduction band. Fermi level in the charge neutral region can be obtain from the neutrality equation $N_d^+ = n_{3D}^o$. Now, from Poisson equation we get:

$$\left(\frac{dV(z)}{dz}\right)^2 = -\frac{2e}{\varepsilon_s}\int_0^V \left[N_d^+(V'(z)) - n_{3D}(V'(z))\right] dV' = \frac{2e}{\varepsilon_s}\Im(V(z)) \qquad (1)$$

Electric field $E(z) = -\frac{dV}{dz}$ can be expressed as $E(z) = \sqrt{\frac{2e}{\varepsilon_s}\Im(V(z))}$. Electric field at the surface can be expressed as:

$$E_s = \sqrt{\frac{2e}{\varepsilon_s}\Im(V_s)} \qquad (2)$$

The 2D negative charge density in the accumulated region is thus $Q_{SC} = -\varepsilon_s E_s$. One can obtain $V_s$ from the equation of charge conservation, which can be expressed as

$$Q_{SS} = -Q_{SC} \qquad (3)$$

When the sample is illuminated with above band gap photons, additional $\Delta n$ density of electrons and the holes are excited in conduction and valence bands, respectively, at the surface. This introduces quasi Fermi levels for electrons ($E_{Fn}$) and holes ($E_{Fh}$). It should be noted that in principle, for both electrons and holes, the quasi Fermi levels vary as a function of $z$ as they gradually moves towards each other and marge with the equilibrium Fermi level at a depth where most of the photons are absorbed as shown schematically in Fig. 5. Since the width of the accumulation region is typically much less than the electron and hole diffusion length, $E_{Fn}$ and $E_{Fh}$ can be considered $z$-independent within the accumulation region. Quasi Fermi levels for electrons and holes just outside the accumulation region can be obtained from $n_{3D}^o + \Delta n = N_c F_{1/2}(E_{Fn} - E_C^o)/k_B T)$ and $\Delta n = N_v F_{1/2}(E_V^o - E_{Fh}/k_B T)$, respectively. Note that $E_C^o$ and $E_V^o$ are the conduction band minimum and valence band maximum, respectively, in the charge neutral region. Considering, $E_{Fn} - E_C^o = \Delta E_{Fn}$ and $E_{Fh} - E_V^o = \Delta E_{Fh}$ to be position independent throughout the accumulation region [18,26,27], the charge density as a function of position [$\rho^*(z)$] can now be expressed as $\rho^*(z) = e[N_d^+(z) + p_{3D}(z) - n_{3D}(z)]$ where, $N_d^+(z) = N_d / 1 + 2\exp[\{E_{Fn} + E_d + eV(z)\}/k_B T]$, $p_{3D}(z) = N_v F_{1/2}[\{-eV(z) - E_{Fh}\}/k_B T]$ and $n_{3D}(z) = N_c F_{1/2}[\{E_{Fn} + eV(z)\}/k_B T]$. Following a similar approach like in case of thermodynamic equilibrium (dark condition), one obtains from Poisson equation, the electric field at the surface in the steady state condition as:

$$E_s^* = \sqrt{\frac{2e}{\varepsilon_s}\int_0^{V_s^*}\left[N_d^+(V'(z)) + p_{3D}(V'(z)) - n_{3D}(V'(z))\right]dV'} = \sqrt{\frac{2e}{\varepsilon_s}\Im^*(V_s)} \qquad (4)$$

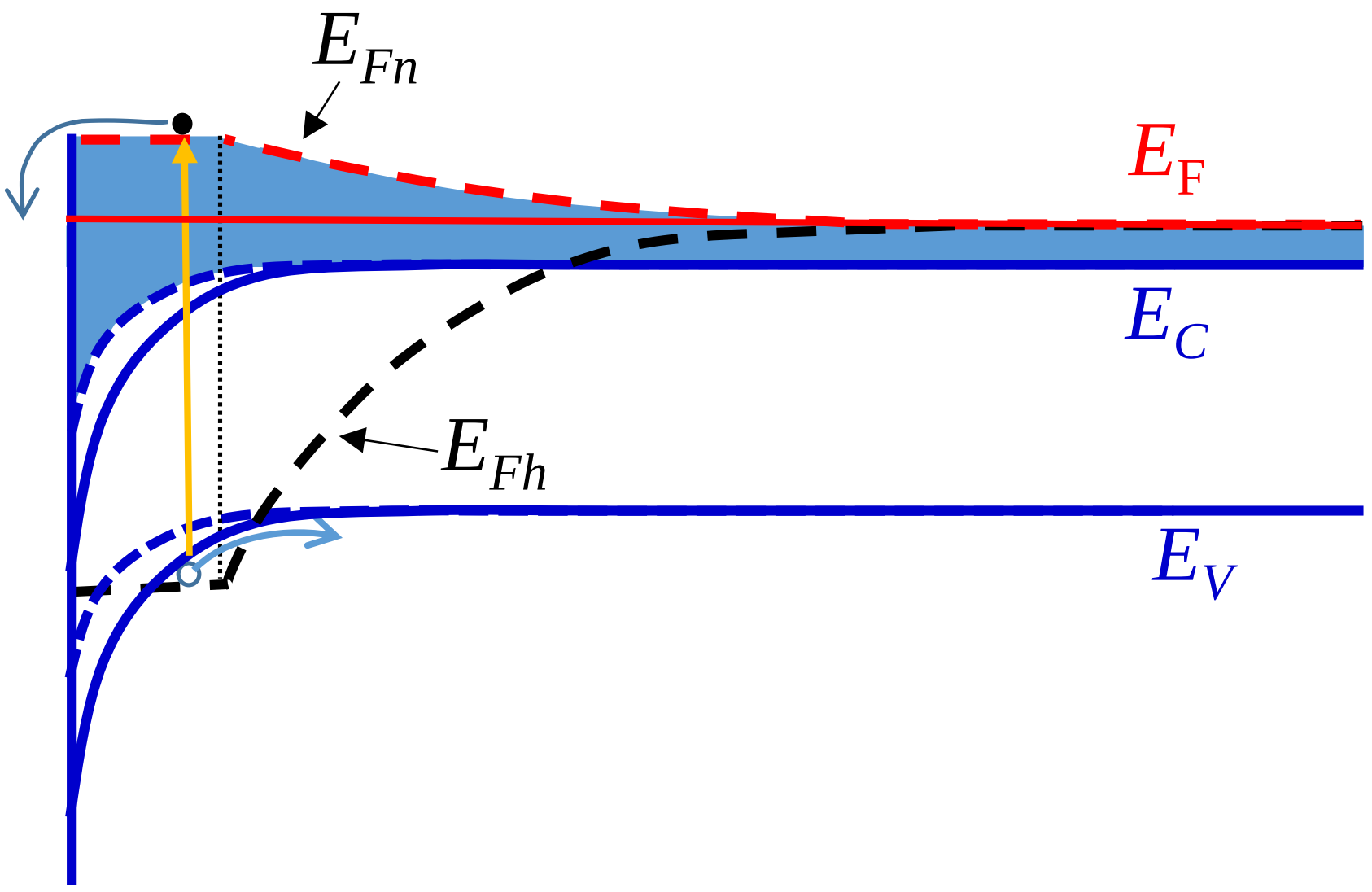

Fig. 5. Schematic depiction of the conduction and valence band edges before (solid blue lines) and after (dashed blue lines) the illumination. Fermi level position in dark (solid red line) and the quasi Fermi level positions for electrons (dashed red line) and holes (dashed black lines) under illumination are also shown schematically.

The charge density of the accumulation layer $Q_{SC}^* = -\varepsilon_s E_s^*$. Surface charge density $Q_{SS}^*$ is modified to

$$Q_{SS}^* = eN_{sd} / \left[1 + \exp\left\{\left(\Delta E_{Fn} + E_{sd} + eV_s^*\right)/k_B T\right\}\right] \qquad (5)$$

$V_s$ can be derived from the charge conservation Eq. (3). We have performed a calculation for a InN layer, which has a background donor concentration of $N_d = 1\times10^{19}$ cm$^{-3}$. Furthermore, the activation energy is considered to be zero for these donors as well as for the donor-like-surfacestates.

Figure 6(a) and (b) show the calculated variation of $eV_s$ and $Q_{SC}$, respectively, as a function of surface donor concentration $N_{sd}$ at $T = 300$ K in dark and illuminated conditions associated with different $\Delta n$. It is interesting to note that for the entire range of surface donor density ($N_{sd}$), $eV_s$ becomes less negative and hence $Q_{SC}$ decreases when illuminated. This means that the band bending at the surface decreases upon illumination [see Fig. 5]. Note that the reduction of surface potential ($e\Delta V_s$) increases with $N_{sd}$. In the inset of Fig. 6(a), the absolute value of $e\Delta V_s$ calculated for $\Delta n = 9\times10^{23}$ m$^{-3}$ is plotted as a function of the magnitude of the surface potential in dark condition $|eV_s^d|$, which shows an increase of $|e\Delta V_s|$ with $|eV_s^d|$.

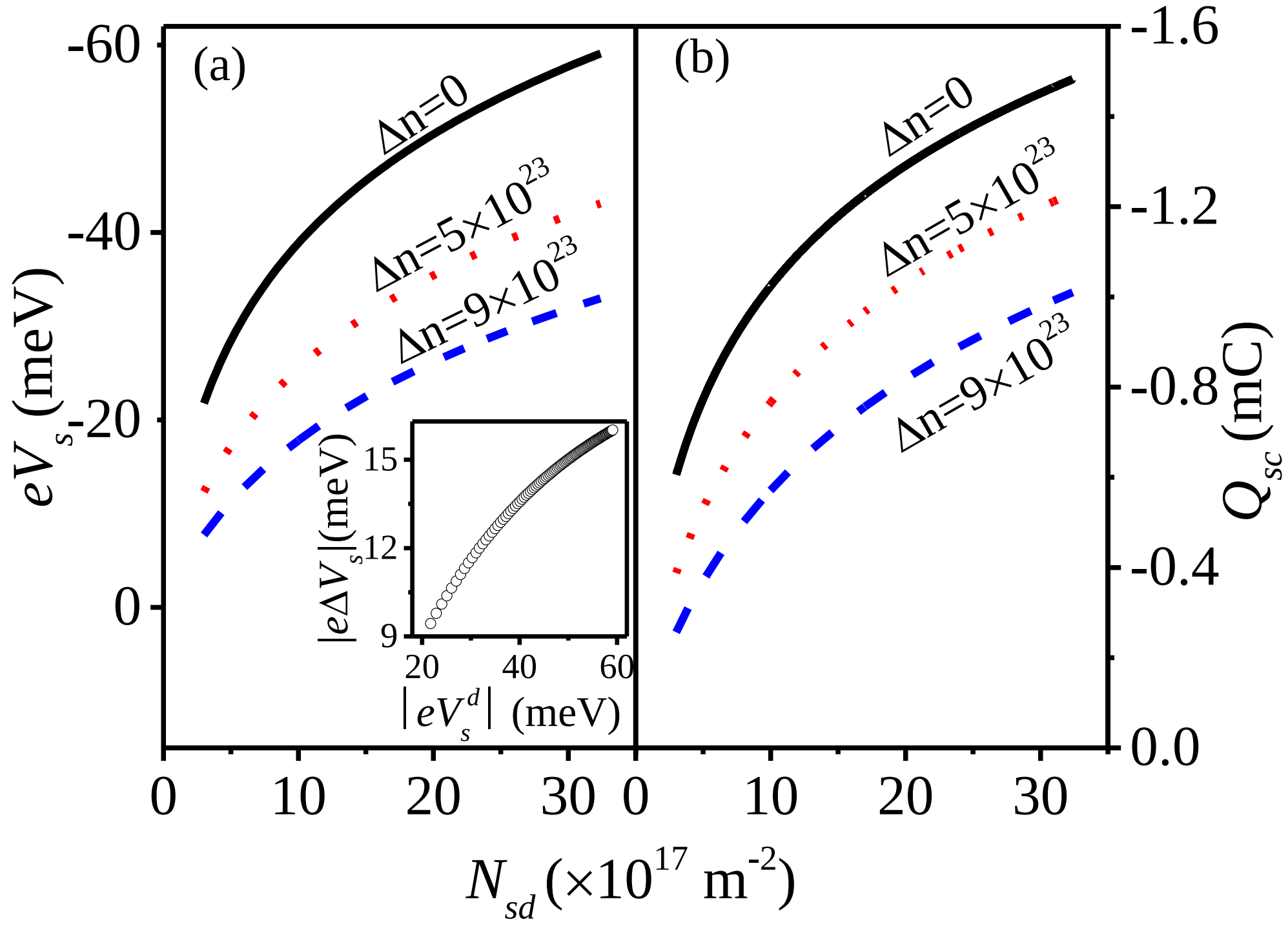

Fig. 6. Variation of the (a) surface potential $eV_s$ and (b) the areal charge density in the accumulation region $Q_{SC}$ as a function of surface donor concentration $N_{sd}$ at $T = 300$ K in dark and illuminated with different $\Delta n$ conditions. Inset shows the variation of the absolute value of the reduction of the surface potential $|e\Delta V_s|$ calculated for $\Delta n = 9\times10^{23}$ m$^{-3}$ as a function of the magnitude of the surface potential in dark condition $|eV_s^d|$.

Fig. 7 depicts the variation of $eV_s$ and $Q_{SC}$ as a function of $\Delta n$ at $T = 300$ K for $N_{sd} = 1\times10^{18}$ m$^{-2}$. Evidently, both the quantities decrease as $\Delta n$ increases. Inset of the figure shows the variation of the electron and hole quasi Fermi levels $E_{Fn}$ and $E_{Fh}$ as a function of $\Delta n$. When illuminated, electron quasi Fermi level moves deeper in the conduction band, while hole Fermi level enters the valence band, as evident from the inset. This upward movement of $E_{Fn}$ results in the reduction in the density of ionized surface donors and hence $Q^*_{SS}$ [Eq. 5]. In order to satisfy the charge conservation [Eq. 3], $Q^*_{SC}$ and thus $|eV_s|$ decreases as shown schematically in Fig.5. This explains the apparently surprising effect of reduction of electron density in the accumulation region upon illumination as shown in Fig.7. However, it should be noted that the conduction band electron concentration in the region away from the 2DEA region increases upon illumination as expected.

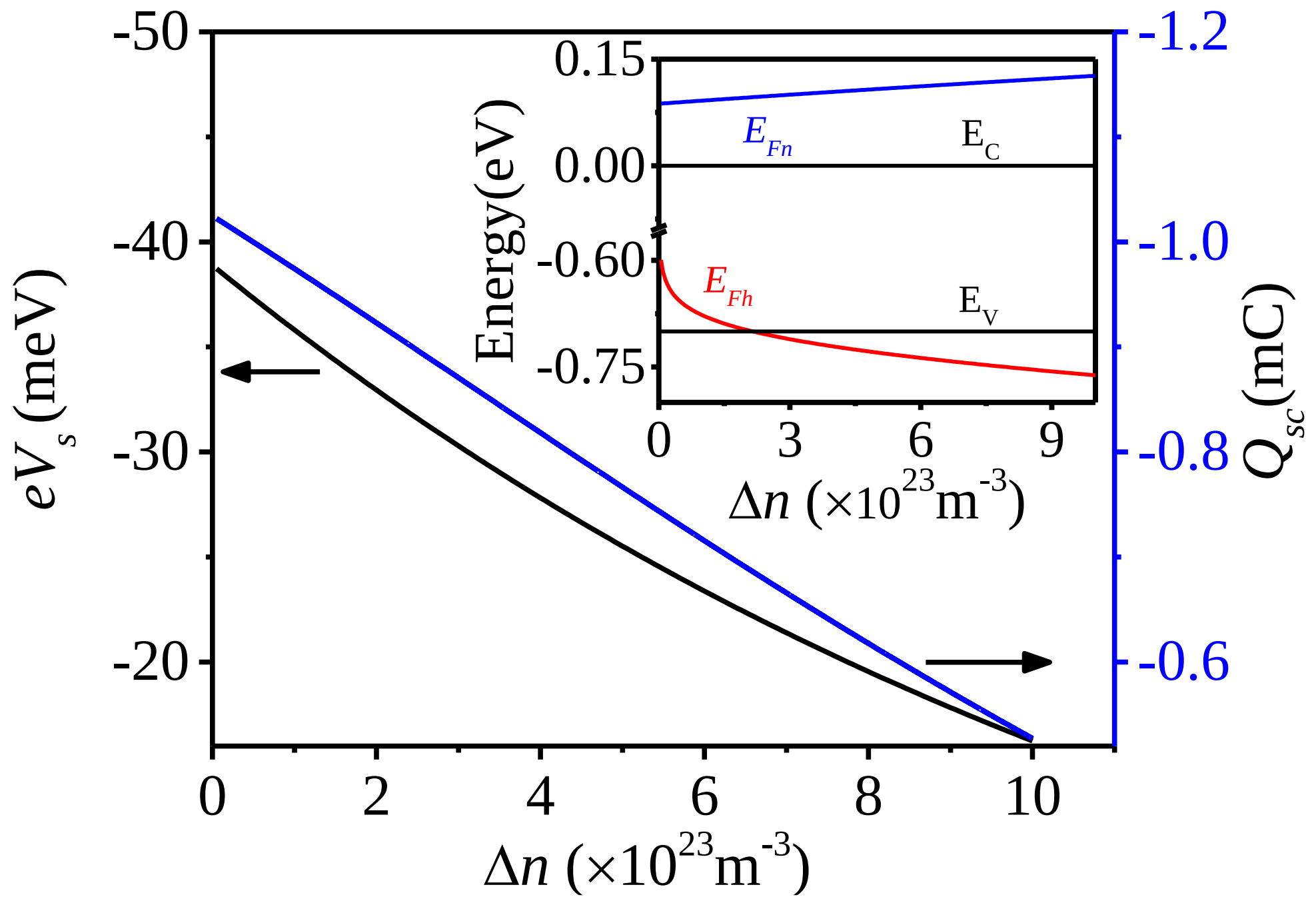

Fig. 7. Variation of $eV_s$ and $Q_{SC}$ as a function of $\Delta n$ at $T = 300$ K for $N_{sd} = 1\times10^{18}$ m$^{-2}$. Inset shows the variation of electron and hole quasi-electron fermi levels under illumination as a function of $\Delta n$.

We believe that the origin of photovoltage developed between different parts of the surface is the inhomogeneity of the surface potential, which is arising due to an inhomogeneous distribution of the surface groups/adatom leading to a nonuniformity in the density of donor-like-surfacestates ( $N_{sd}$ ). It is plausible that the inhomogeneity in the surface adsorption of these groups/adatoms is resulting from a nonuniform distribution of surface roughness. If there is a spatial variation of $N_{sd}(\vec{r})$, where $\vec{r} = x\hat{x} + y\hat{y}$ a two dimensional position vector on the surface, one can expect a variation of both $eV_s(\vec{r})$ and $Q_{SC}(\vec{r})$ over the surface. This can lead to a surface current density distribution $\vec{K}_n(\vec{r})$ for the electrons in the 2DEA region, which can be expressed as

$$\vec{K}_n(\vec{r}) = -e\mu_n Q_{SC} \vec{\nabla}_r V_s + D_n \vec{\nabla}_r Q_{SC} \qquad (6)$$

Where, $\vec{\nabla}_r$ is the 2D gradient operator, $\mu_n$ and $D_n$ are the mobility and diffusion coefficient of the electrons in the accumulation region. In thermal equilibrium(dark condition), the drift and diffusion current balance each other in such a way that the net current density $\vec{K}_n = 0$ maintained everywhere, which leads to a relationship between $\mu_n$ and $D_n$. However, under illumination condition, $\vec{K}_n \neq 0$ anymore, rather $\vec{K}_n$ becomes a function of $\vec{r}$ through Eq. (6). It should be mentioned that a similar equation like Eq. (6) can be written for hole current density $\vec{K}_p$. But, $\vec{K}_p$ can be neglected as compared to $\vec{K}_n$ as $\mu_p$ $(D_p)$ is expected to be much less than $\mu_n$ $(D_n)$. Let us consider two contact pads A and B are separated by a distance of $l$ along x-axis. Furthermore, suppose that $N_{ds}$ varies along x-direction as $N_{ds}(x) = N_o(1+\theta x)$, where $\theta$ and $N_o$ are constants. In the illuminated condition, surface band bending and hence the electron concentration in the accumulated region varies from A to B as shown schematically in Fig. 8. This leads to a surface current density $K_n(x)$ flowing between the contacts. An EMF developed between A and B can be estimated as

$$\xi_{AB} = \int_A^B E_x(x)dx = \int_A^B \frac{K_n(x)}{\sigma_{al} w} dx \qquad (7)$$

Where, $\sigma_{al}$ and $w$ the electrical conductivity and width of the accumulation layer.

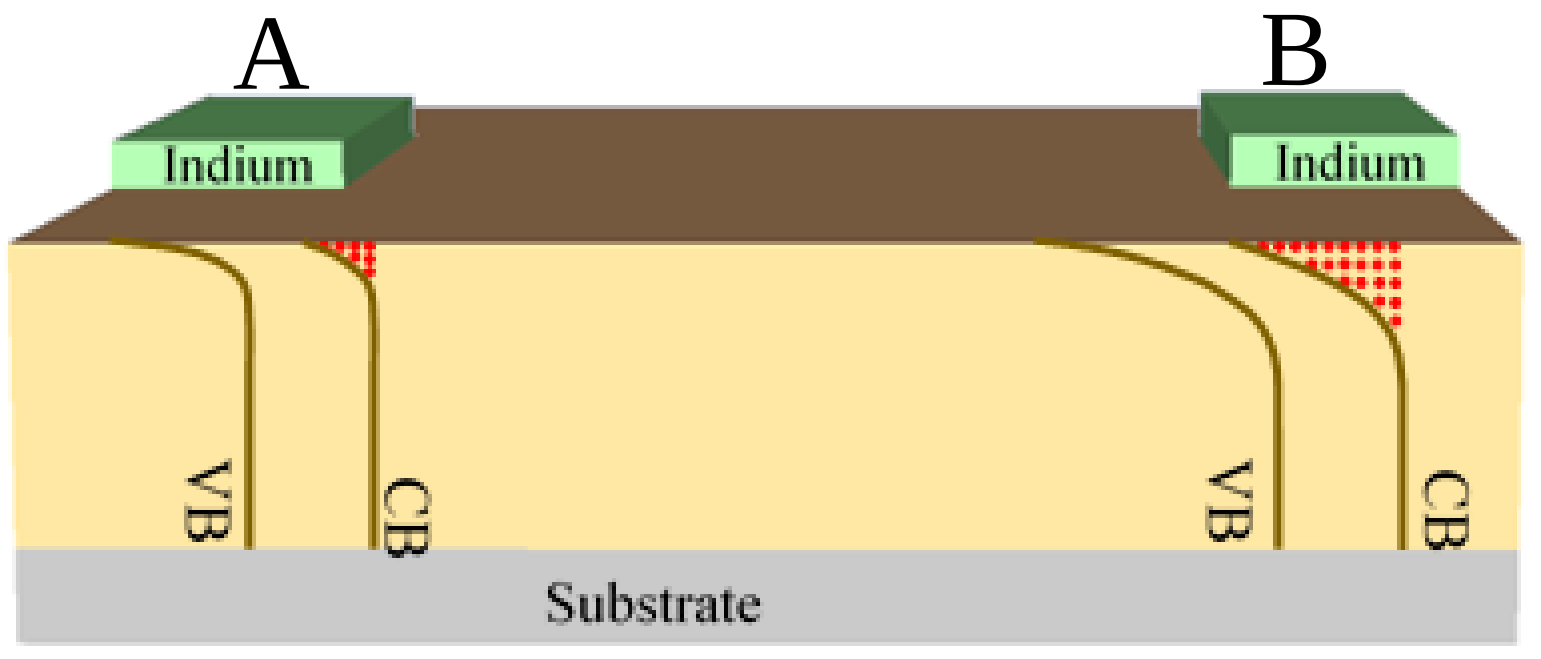

Fig. 8. Schematic representation of band bending at two different locations. Differences in the surface potentials $eV_s$ and the 2D accumulated electron at the surface $Q_{SC}$ are also portrayed.

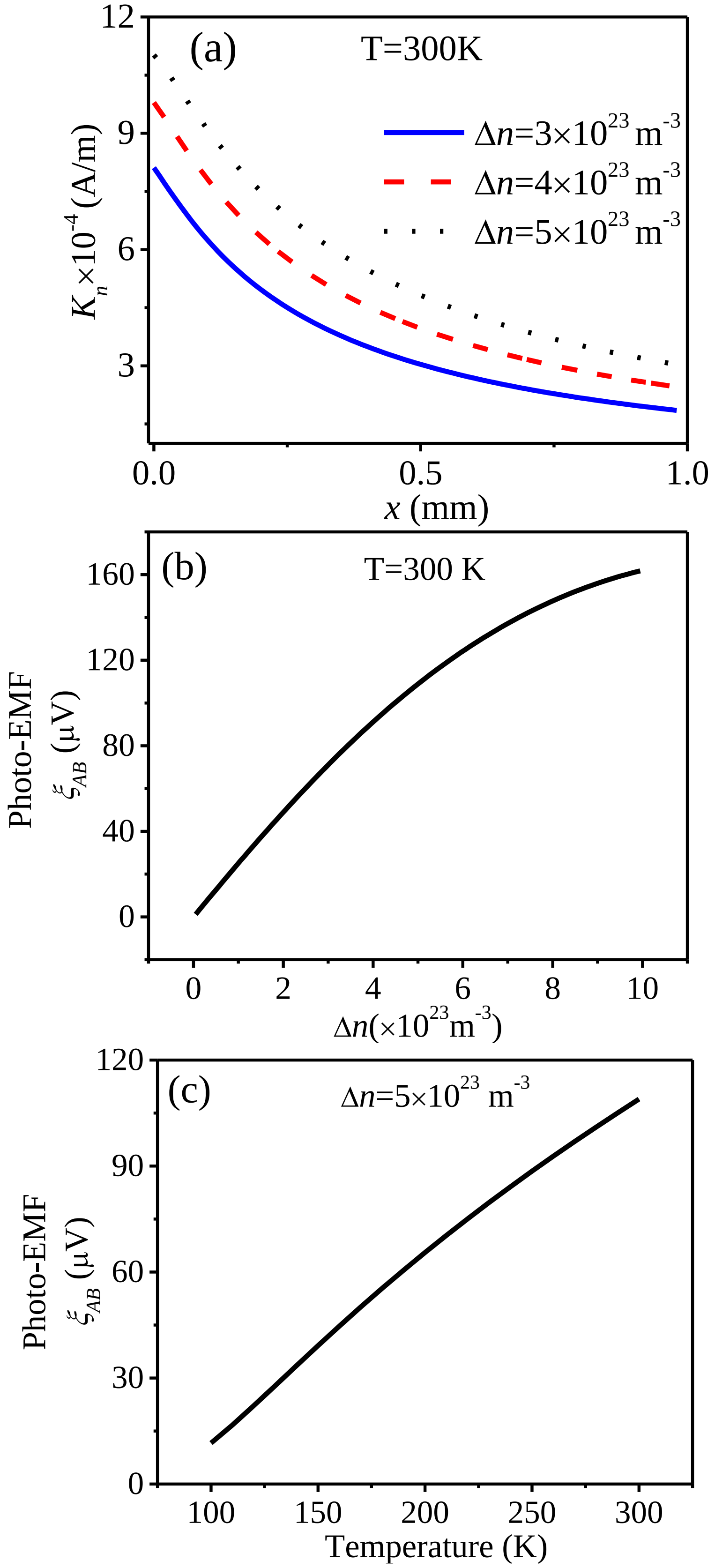

Fig. 9. (a) Surface photo-induced surface current density $K_n(x)$ as a function of distance $x$ when illuminated with different values of $\Delta n$. (b) Variation of photo-EMF developed between contacts A and B $\xi_{AB}$ as a function of $\Delta n$. (c) Variation of $\xi_{AB}$ as a function of temperature.

Figure 9 represents the results of our calculation for the surface current $K_n$ and EMF developed between A and B ($\xi_{AB}$) considering $\theta = 10^4$ m$^{-1}$ and $N_o = 3\times10^{17}$ m$^{-2}$ and the resistance of the accumulation region $(\sigma_{al} w)^{-1}$ is considered to be 200 ohm. Fig. 9(a) shows $K_n$ as a function of the distance $x$ from the contact A for different values of $\Delta n$. Clearly, $K_n$ for all cases increases as the excitation power is enhanced. $\xi_{AB}$ is plotted as a function of $\Delta n$ in Fig. 9(b). Evidently, the calculated values of $\xi_{AB}$ are of the same order of magnitude as the values of $V_{ph}^{max}$ observed in Fig. 1 and 2. Moreover, $\xi_{AB}$ shows an enhancement with the excitation power. Note that the magnitude of the photovoltage build-up in our sample has been found to increase with the excitation power by about the same factor that is theoretically predicted as shown in Fig.2(b). Fig. 9(c) depicts the variation of $\xi_{AB}$ as a function of temperature $T$ for $\Delta n = 5\times10^{23}$m$^{-3}$. Enhancement of $\xi_{AB}$ with $T$ is again consistent with the experimental observation of Fig.2(c). These findings strongly validate our model.

## E. Surface potential inhomogeneity: Kelvin probe force microscopy

Note that our theory attributes the photo-EMF effect observed in this system [see Fig. 1 and 2] to the inhomogeneous distribution of certain groups/adatoms attached on the surface leading to the formation of donor-like-surfacestates. A spatial inhomogeneity of the attachment of these groups/adatoms with the surface can give rise to an inhomogeneity in the surface band bending and hence in the depth of the electron confining potential (surface potential $V_s$), which is the cause for the photo-EMF effect. However, the identification of those groups/adatoms could not be made. One plausible reason is oxygen, which might be chemisorbed on the layer surface once the sample is taken out of the growth chamber. Probably, such attachments depend on the surface roughness. Inhomogeneity in surface roughness might be resulting in a nonuniform distribution of oxygen adatoms and hence in potential variation over the surface. It should be noted that the band bending observed in c-GaN epitaxial layers has indeed been attributed to surface oxygen [28]. It is noteworthy that in Fig.1(b), the photo-EMF is found to be less in humid environment or in ambient condition as compared to that is recorded under vacuum. This might suggest that the attachment of additional groups (such as OH-groups) on the surface reduces the density of the donor-like-surfacestates. The study of identification of the surface attached species leading to the downward band bending at the surface is currently underway.

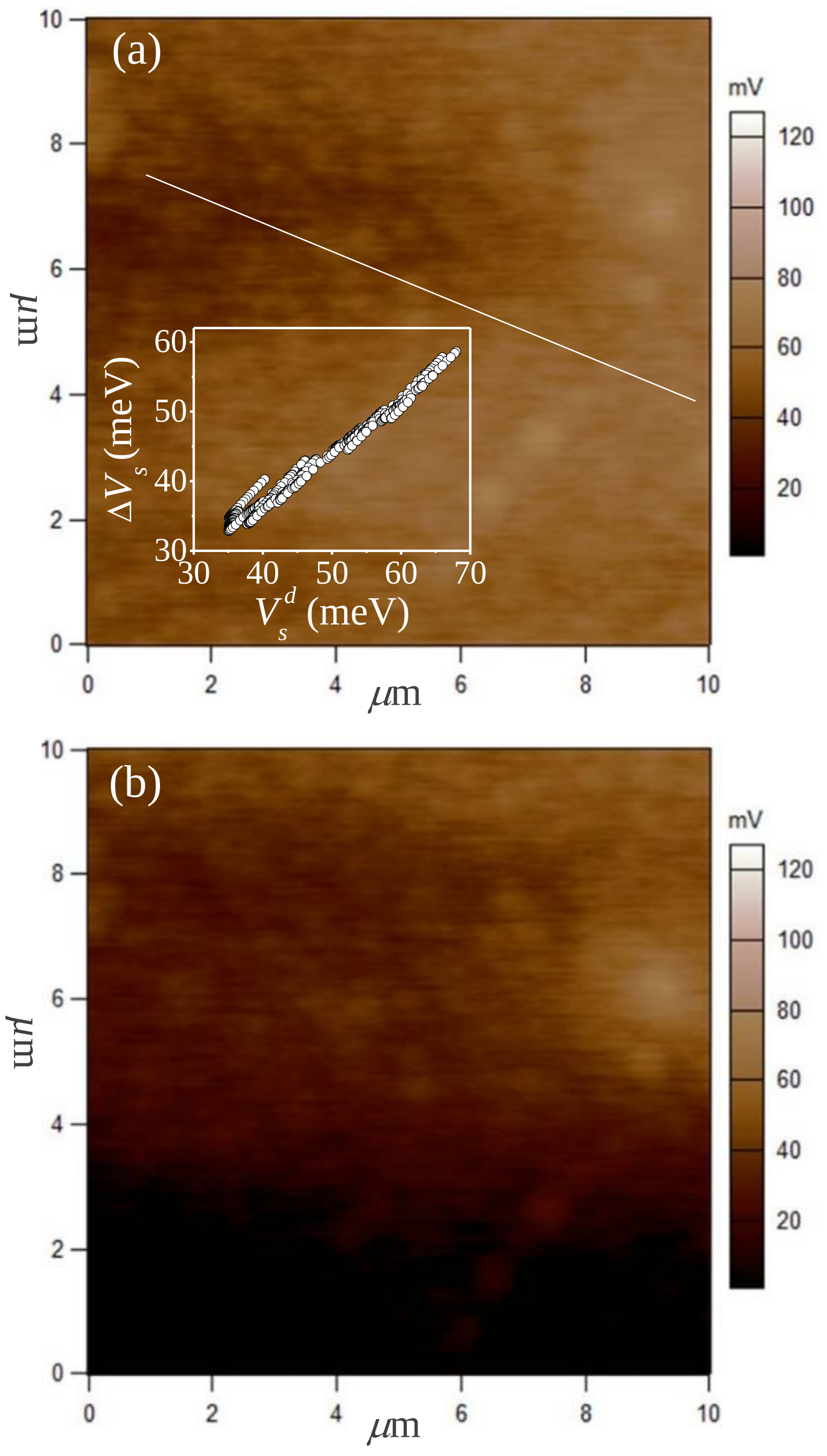

Fig. 10. Kelvin probe force microscope images under (a) dark and (b) illumination conditions. In the inset of (a) the change in surface potential $\Delta V_s$ upon illumination is plotted as a function of the surface potential in dark condition $V_s^d$ . The line, along which these data were scanned, is also marked in the image.

Here, we have used Kelvin probe force microscopy (KPFM) to examine whether inhomogeneity in the distribution of the potential at the surface of the InN layer really exists or not. Fig. 10 compares the KPFM images recorded on a portion of the sample surface before [Fig.10(a)] and after [Fig.10(b)] illumination with white light. It is noticeable that the contact potential difference (CPD) remains to be positive all over the area both under dark and illuminated conditions but the magnitude of CPD decreases upon illumination. This general tendency of positive CPD and its reduction under light exposure has been found to be followed in every spot selected randomly on the sample surface. Observation of positive CPD implies negative potential all over the surface. This is indeed consistent with the accumulation of electron on the surface and corroborated by the observations of our XPS study in Fig.4. Furthermore, the reduction of the magnitude of the surface potential upon illumination is in fact congruous with our theoretical predictions shown in Fig. 6 and 7. Change in surface potential $|e\Delta V_s|$ upon illumination is plotted as a function of the surface potential in dark condition $|eV_s^d|$ in the inset of Fig. 10(a). The line, along which these data were scanned, is also marked in the image. Interestingly, all data points fall on a line, implying a systematic variation of $|e\Delta V_s|$ with $|eV_s^d|$. Note that we have recorded $e\Delta V_s$ versus $eV_s^d$ data from line scans taken at several randomly chosen locations on these images as well as on similar images taken from other parts of the surface. In all cases, $|e\Delta V_s|$ has been found to increase consistently with $|eV_s^d|$.

This observation is indeed in accordance with that is theoretically predicted in the inset of Fig. 6(a). Finally, variation of surface potential over a length scale of a few microns can clearly be seen in both the images of Fig. 10. In order to know whether any spatial inhomogeneity of $V_s$ with longer length scale exist on the surface or not, we have taken $1\mu m \times 1\mu m$ KPFM images at several locations laying on the line joining contact pads 2 and 3. Average $V_s$ for each of the images is plotted as a function of its distance from the contact pad 2 in Fig. 11. It is evident from the figure that $V_s$ steadily decreases as one proceeds from contact pads 2 to 3. This observation clearly provides a strong justification to our argument of attributing photovoltage effect to the inhomogeneity in surface potential distribution.

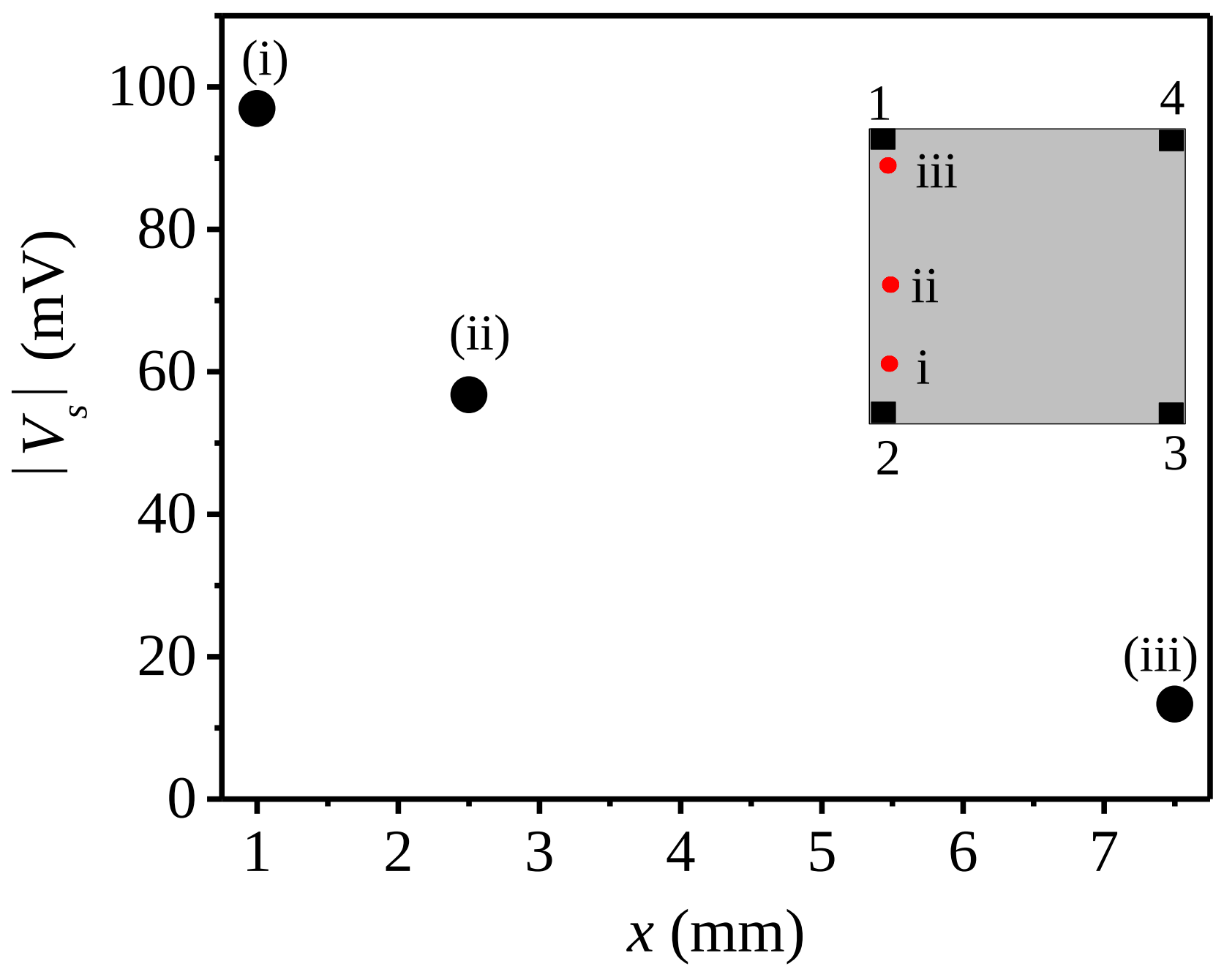

Fig. 11. Average $V_s$ at different locations as shown in the inset.

## F. Photo-EMF Device Proposal

Conventional photo-voltaic devices, where the EMF is developed across the thickness of the layer, requires the device to be grown on a conducting substrate. In contrast, the surface photo-EMF effect is entirely a surface phenomenon, it can be exploited by depositing surface contacts at the two sides of a density gradient of the donor-like-surfacestates ( $N_{sd}$ ) even though the layer is deposited on top of an insulating substrate. One can envisage to fabricate a large number of photo-EMF cells connected in series out of a single layer. One such device proposal is depicted in Fig. 12. A $N_{sd}$ gradient along $x$-direction can be artificially introduced on the surface of a InN film in the form of stripes as presented in the figure. One possible way to introduce such a variation of $N_{sd}$ could be to cover the surface with a photoresist having a gradient of thickness along that direction. The covered surface can then be exposed to a plasma. The varied thickness of the protective layer will lead to different degrees of surface modification, resulting in a gradient of $N_{sd}$ along that direction. On each stripe, an array of metal contacts is deposited and then these pads are pair-wise isolated through a suitable lithography and lift-off process. When illuminated, each pair of contacts can now act as a battery. These batteries can be connected in series by depositing metallic interconnects as shown in the figure. If we consider a cell, where two 1 mm$\times$1 mm contact pads are separated by 1 mm, the short circuit current $I_{SC} = K_n \times 1$ mm and open circuit voltage $V_{OC} = \xi_{AB}$ can be estimated as $\approx 10^{-6}$ A and $\approx 10^{-4}$ V, respectively, from the results of our calculation shown in Fig. 9. Therefore, the maximum theoretical power output from such a device $P_{\max}^{th} \approx 0.1$ nW. Note that for the calculation of $K_n$ and $\xi_{AB}$ in Fig. 9, the gradient of $N_{sd}$ is considered to be $\theta = 10^4$ m$^{-1}$ and the resistance of the accumulation region $(\sigma_{al} w)^{-1}$ in Eq.7 is considered to be 200 ohm, which is typically the case for these samples. Here, the estimation of the maximum output power is made for $\Delta n = 5\times10^{23}$ m$^{-3}$. Considering an absorption coefficient $\alpha = 10^5$ cm$^{-1}$ of InN [1] for the photon energy of 2 eV and noting that the thickness of the accumulation layer is $w \approx 2$ nm, one can make an estimate about the power of the photons needed to generate $\Delta n_{SC} = w\Delta n = 10^{15}$ m$^{-2}$ areal density of electrons in the accumulation region. If we consider that each photon generate one electron-hole pair, then $G = \Delta n_{SC} / \tau$ should be the rate of absorption of photons per unit area of the accumulation region. Note that only $(1-e^{-\alpha w})$ fraction of the incident photon energy flux $I_o$ is absorbed in the accumulation region. Therefore, $I_o = \hbar\omega\Delta n_{SC} / \tau(1-e^{-\alpha w})$, where $\hbar\omega$ the incident photon energy. From the time-dynamics of the photo-EMF profiles shown in Fig. 1, we get an estimate of $\tau \approx 10$ sec., which leads to $I_o \approx 2\times10^{-3}$ Wm$^{-2}$, meaning the incident photon power falling on the device area has to be $\approx 2$ nW, in order to get a power output of $P_{\max}^{th} \approx 0.1$ nW from a single cell. The efficiency $\eta$ can thus be estimated to be 5%. It is worthy to note that $P_{\max}^{th}$ can be increased by several factors through the enhancement of $\theta$. For an example, we have calculated $P_{\max}^{th} \approx 1$ nW and hence $\eta \approx 50\%$ for $\theta = 10^5$ m$^{-1}$.

In order to highlight further the application perspective of the effect, we have performed photocurrent versus voltage measurements between a pair of indium contacts (3 mm apart) deposited on the surface of an InN layer. The sample is illuminated with a 35 Watt tungsten-halogen lamp. The results are shown in Fig. 13. The output power $P_{out} = IV$ is also plotted as a function of the applied voltage in the same figure. Evidently, the maximum experimental output power $P_{\max}^{ex}$ has been found to be ~0.3 nW. While the fill factor $FF = P_{\max}^{ex} / I_{SC}V_{OC}$ with $I_{SC}$, the short circuit current and $V_{OC}$, the open circuit voltage is estimated to be 0.5. Note that a similar order of magnitude of the output power ($P_{\max}^{th} \approx 0.1$ nW) is estimated theoretically in our model calculation.

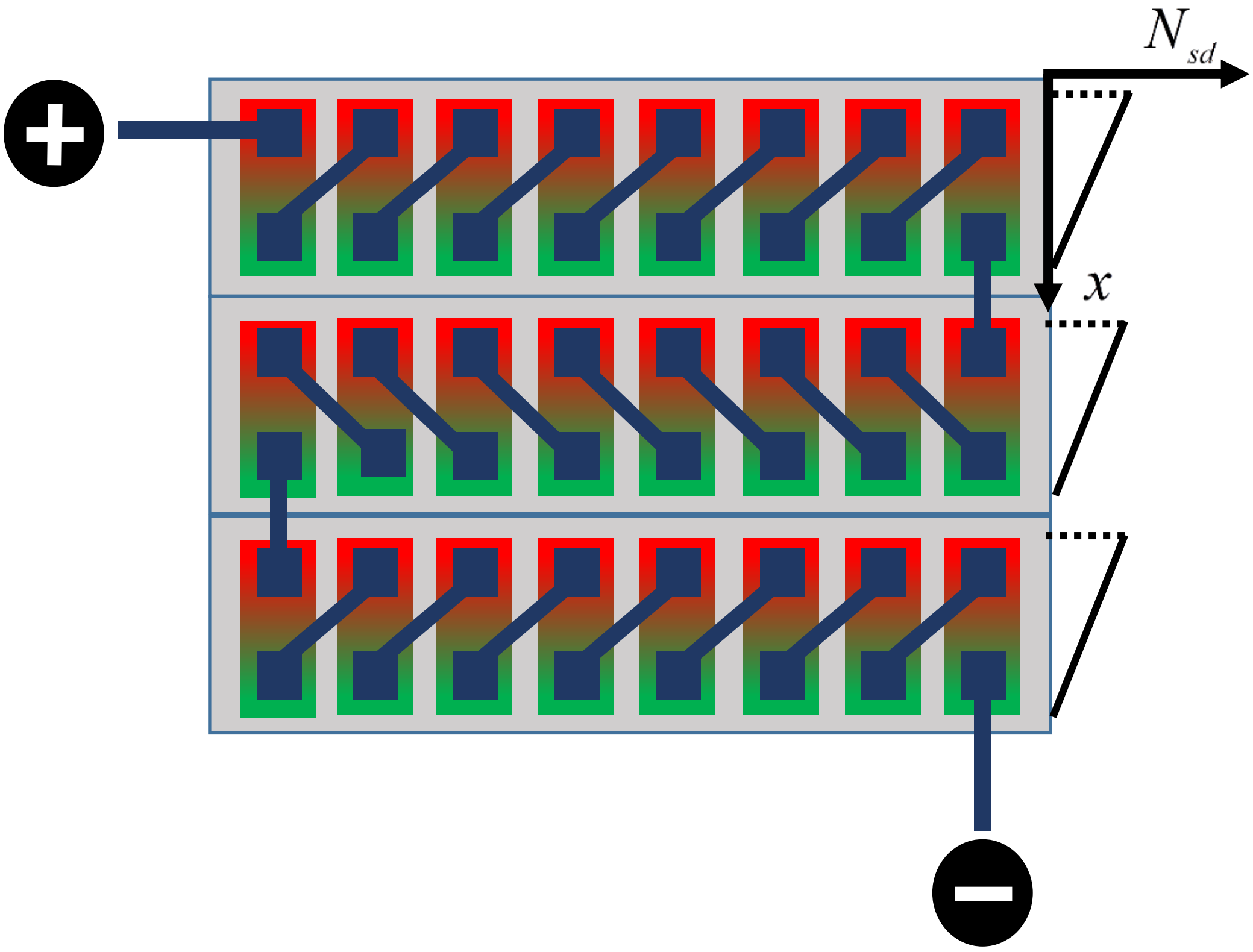

Fig. 12. Schematic depiction of a proposed device: An array of photo-EMF cells connected in series. Red and green contrast represent the gradient of $N_{sd}$ .

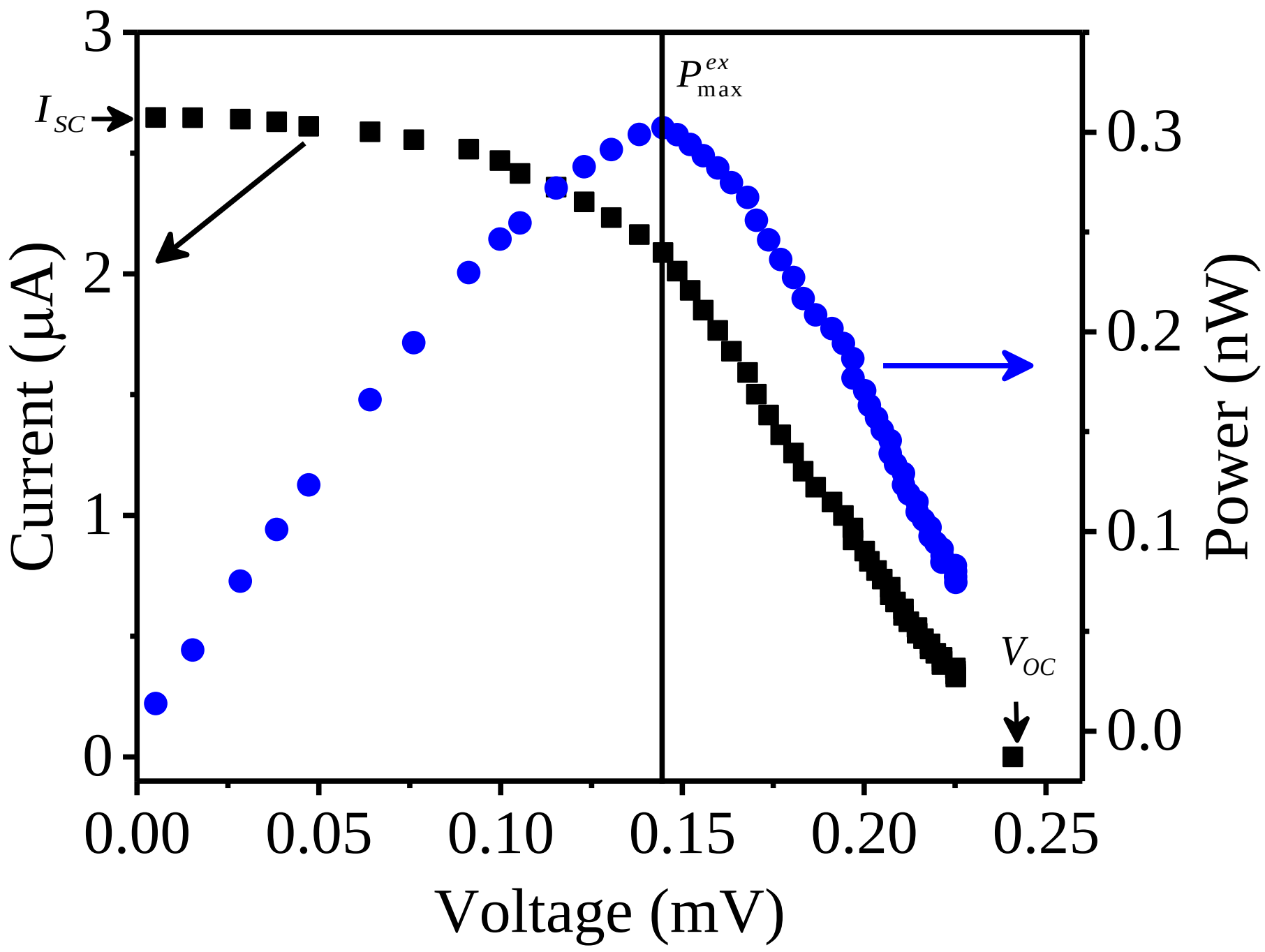

Fig. 13. Photocurrent versus voltage characteristic between two indium contacts deposited on the surface of an InN epitaxial layer (black solid squires). Positions of $I_{SC}$ and $V_{OC}$ are indicated by arrows. Output power as a function of voltage (Blue solid circles).

### G. Understanding Negative Photoconductivity Effect

It should be noted that our model predicts a reduction of the magnitude of the surface potential $V_s$ and hence in the concentration of the accumulated 2D surface electrons upon illumination as shown in Fig. 6 and 7. This prediction is corroborated by the observations of the KPFM shown in Fig. 10. It can be understood that the conductance of these layers should be governed by two parallel conduction channels; a 2D electron accumulation region on the surface and the underlying bulk region. Because of quantum confinement effect, electron mobility in the 2D channel is expected to be substantially more than that is in the bulk region. Since the electron concentration in the 2D channel is decreased upon illumination, the overall conductance could be reduced even though the carrier concentration in the bulk region increases. This might explain the negative photoconductivity effect observed in these layers [9,16,17].

It should be noted that according to the model, the main reason for the observation of photo-EMF as well as negative photoconductivity effects in this material is the formation of 2D electron accumulation (2DEA) region. Such a region is formed in order to accommodate the excess electrons transferred from certain donor-like-surfacestates likely to be resulting due to the adsorption of certain groups/adatoms on the surface. This thin two dimensional electron gas (2DEG) channel can have a much higher mobility as compared to the underlying bulk layer as a result of the suppression of scattering cross-section due to reduced dimensionality and hence can dominate the conducting property of the material. In addition, a non-uniformity in the distribution of donor-like-surfacestates and hence in the surface potential is necessary to give rise to the photo-EMF effect. It is noteworthy that the surface band bending in usual semiconductors such as n-type GaAs, GaN, ZnO etc. is found to result in the formation of a positive depletion region (upward band bending) instead of a negative 2DEA region at the surface [28-30]. A depletion region, where the charges are stationary, cannot give rise to these effects. This might explain why both the effects are so unique to n-type InN epitaxial layers.

## IV. Conclusion

*c*-oriented InN epitaxial layers grown on c-sapphire substrates using chemical vapour deposition (CVD) technique, exhibit surface photo-EMF effect, in which an EMF is developed between different parts of the sample surface when it is illuminated with above band gap light. It has been found that while the polarity of this surface photo-EMF depends on the location of the contacts, its magnitude is influenced by the temperature, excitation power and the environmental conditions. The effect is notably different from surface photovoltaic effect, in which a voltage is developed between the surface and the bottom of the layer. Since InN has a band gap of only about 0.7 eV, this phenomenon could potentially be useful for photovoltaic application. The material also shows negative photoconductivity effect, coexistence of which with the surface photo-EMF effect results in a unique photoconductivity response, where the magnitude as well as the sign of the photoconductivity response change as a function of the applied bias between the contacts. A theoretical model is proposed, where the formation of the 2D electron gas (2DEG) region just below the film surface is attributed to certain donor-like-surfacestates, which are likely to be resulting due to the adsorption of certain groups/adatoms on the surface. According to the model, the photo-EMF effect can be explained in terms of a spatially inhomogeneous distribution of these species resulting in a lateral non-uniformity of depth of the potential profile ($V_s$) confining the 2DEG. Existence of an inhomogeneous distribution of $V_s$ over the film surface, has indeed been found in KPFM studies carried out on these layers.

## Acknowledgements:

We acknowledge Sophisticated Analytical Instrument Facility (SAIF) and central facilities of IIT Bombay, Centre of excellence for nanoelectronics (CEN) of IIT Bombay for providing various experimental facilities. This work was supported by Department of Science and Technology (DST) under Grant No: SR/S2/CMP–71/2012 and Council of Scientific & Industrial Research (CSIR) under Grant No: 03(1293)/13/EMR-II, Government of India.

## Reference:

1. T. L. Tansley and C. P. Foley, Optical band gap of indium nitride, J. Appl. Phys. 59, 3241 (1986).
2. A. Knübel, R. Aidam, V. Cimalla, L. Kirste, M. Baeumler, C. C. Leancu, V. Lebedev, J. Wallauer, M. Walther and J. Wagner, Transport characteristics of indium nitride (InN) films grown by plasma assisted molecular beam epitaxy (PAMBE), Phys. Status Solidi C, 6, 1480 (2009).
3. Z. H. Zhang, W. Liu, Z. Ju, S. T. Tan, Y. Ji, Z. Kyaw, X. Zhang, L. Wang, X. W. Sun and H. V. Demir, InGaN/GaN multiple-quantum-well light-emitting diodes with a grading InN composition suppressing the Auger recombination, Appl. Phys. Lett., 105, 033506 (2014).

4. C. Rivera, J. Pereiro, Á. Navarro, E. Muñoz, O. Brandt and H. T. Grahn, Advances in Group-III-Nitride Photodetectors, TOEEJ, 4, 1 (2010).
5. D. V. P. McLaughlin and J. M. Pearce, Progress in Indium Gallium Nitride Materials for Solar Photovoltaic Energy Conversion, Metall. Mat. Trans. A, 44, 1947 (2013).
6. S. N. Mohammad and H. Morkoc, Progress and prospects of group-III nitride semiconductors, Prog. Quantum Electron. 20, 361 (1996).
7. V. M. Polyakov and F. Schwierz, Low-field electron mobility in wurtzite InN, Appl. Phys. Lett., 88, 032101 (2006).
8. C. G. Van de Walle, J. L. Lyons and A. Janotti, Controlling the conductivity of InN, physica status solidi (a) 207, 1024 (2010).
9. P. C. Wei, S. Chattopadhyay, M. D. Yang, S. C. Tong, J. L. Shen, C. Y. Lu, H. C. Shih, L. C. Chen and K. H. Chen, Room-temperature negative photoconductivity in degenerate InN thin films with a supergap excitation, Phys. Rev. B, 81, 045306 (2010).
10. T. Inushima, Superconductivity of InN caused by In-In nano-structure, phys. stat. sol. (c) 4, 660 (2007).
11. G. J. Papaioannou, M. Nowak and P. C. Euthymiou, Influence of illumination intensity on negative photoconductivity of Si ion-implanted GaAs:Cr, J. Appl. Phys., 65, 4864 (1989).
12. R. Sreekumar, R. Jayakrishnan, C. Sudha Kartha and K. P. Vijayakumar, Anomalous photoconductivity in gamma $In_2Se_3$, J. Appl. Phys., 100, 033707 (2006).
13. Y. Han, X. Zheng, M. Fu, D. Pan, X. Li, Y. Guo, J. Zhao and Q. Chen, Negative photoconductivity of InAs nanowires, Phys. Chem. Chem. Phys., 18, 818 (2016).
14. V. T. Igumenov, D. A. Kichigin, O. A. Mironov and S. V. Chistiakov, Nonequilibrium galvanomagnetic effects of quasi-2D electrons in p-InSb/i-GaAs heteroepitaxial structures, JETP Letters, 38, 459-463 (1983).
15. M. J. Chou, D. C. Tsui and G. Weimann, Negative photoconductivity of two-dimensional holes in GaAs/AlGaAs heterojunctions, Appl. Phys. Lett., 47, 609 (1985).
16. L. Guo, X. Wang, L. Feng, X. Zheng, G. Chen, X. Yang, F. Xu, N. Tang, L. Lu, W. Ge and B. Shen, Temperature sensitive photoconductivity observed in InN layers, Appl. Phys. Lett., 102,072103 (2013).
17. L. Guo, X.Q. Wang, X.T. Zheng, X.L. Yang, F.J. Xu, N. Tang, L.W. Lu, W.K. Ge, B. Shen, L.H. Dmowski and T. Suski, Revealing of the transition from n- to p-type conduction of InN:Mg by photoconductivity effect measurement, Sci. Rep., 4, 4371 (2014).

18. L. Kronik and Y. Shapira, Surface photovoltage phenomena: theory, experiment, and applications, Surf. Sci. Rep. 37, 5 (1999)
19. H. Dember, Photoelectromotive force in cuprous oxide crystals, Physik. Z. 32, 554 (1931); H. Dember, A crystal photocell, Physik. Z. 32, 856 (1931).
20. B. K. Barick, N. Prasad, R. K. Saroj and S. Dhar, Structural and electronic properties of InN epitaxial layer grown on c-plane sapphire by chemical vapor deposition technique, J. Vac. Sci. Technol. A, 34, 051503 (2016).
21. K. A. Rickert, A. B. Ellis, F. J. Himpsel, H. Lu, W. Schaff, J. M. Redwing, F. Dwikusuma and T. F. Kuech, X-ray photoemission spectroscopic investigation of surface treatments, metal deposition, and electron accumulation on InN, Appl. Phys. Lett., 82, 3254 (2003).
22. B. K. Barick, C. Rodríguez-Fernández, A. Cantarero and S. Dhar, Structural and electronic properties of InN nanowire network grown by vapor-liquid-solid method, AIP Adv 5, 057162 (2015).
23. K. S. A. Butchera, A. J. Fernandes, P. P. T. Chen, M. Wintrebert-Fouquet, H. Timmers, S. K. Shrestha, H. Hirshy, R. M. Perks and B. F. Usher, The nature of nitrogen related point defects in common forms of InN, J. Appl. Phys., 101, 123702 (2007).
24. H. Wang, D. S. Jiang, L. L. Wang, X. Sun, W. B. Liu, D. G. Zhao, J. J. Zhu, Z. S. Liu, Y. T. Wang, S. M. Zhang and H. Yang, Investigation on the structural origin of n-type conductivity in InN films, J. Phys. D Appl. Phys., 41,135403 (2008).
25. E. A. Davis, S. F. J. Cox, R. L. Lichti and C. G. Van de Walle, Shallow donor state of hydrogen in indium nitride, Appl. Phys. Lett., 82, 592 (2003).
26. C. G. B. Garrett and W. H. Brattain, Physical Theory of Semiconductor Surfaces, Phys. Rev., 99, 376 (1955).
27. A. Many, Y. Goldstein and N. B. Grover, *Semiconductor Surfaces*, 2nd ed., (North-Holland, Amsterdam, 1971).
28. M. A. Garcia, S. D. Wolter, T. H. Kim, S. Choi, J. Baier and A. Brown, Surface oxide relationships to band bending in GaN, Appl. Phys. Lett., 88, 013506, (2006).
29. Z. W. Deng, R. W. M Kwok, W. M. Lau, L. L. Cao, Time-resolved measurement of surface band bending of cleaved GaAs(110) and InP(110) by high resolution XPS, Appl. Surf. Sci., 158, 58 (2000).

30. Y. J. Lin and C. L. Tsai, Changes in surface band bending, surface work function, and sheet resistance of undoped ZnO films due to $(NH_4)_2S_x(NH_4)_2S_x$ treatment, J. Appl. Phys. 100, 113721 (2006)